\documentclass[twocolumn]{aastex62}

\usepackage{amsmath,amssymb,amstext}

\usepackage[all]{hypcap}
\def\baselinestretch{0.96}
\usepackage{graphicx}
\usepackage{subfigure}
\usepackage{cleveref}
\usepackage{natbib}

\crefformat{section}{\S#2#1#3}

\newcommand{\src}[1]{4U~1907+09}
\newcommand{\as}{AstroSat}
\newcommand{\lx}{LAXPC}

\newcommand{\neh}{NEWHCUT}

\newcommand{\ecyc}[1]{\ensuremath{E_{\rm{C}}}}
\newcommand{\crsfeq}[1]{\ensuremath{
   \ecyc{#1}= 11.6\frac{B}{10^{12}\,\rm{G}}(1 + z)^{-1}
}}

\begin{document}
\title{ \bf Pulse Phase Variation of Cyclotron Line in HMXB 4U 1907+09 with ASTROSAT LAXPC}
\author{Varun}
\affil{{\small Raman Research Institute, C V Raman Avenue, Sadashivanagar, Bangalore 560080, India}}

\author{Pragati Pradhan}
\affil{{\small Department of Astronomy and Astrophysics, Pennsylvania State University, Pennnsylvania, 16802, US}}
\affil{{\small St Joseph's College, Singmari, Darjeeling, 734104, India}}

\author{Chandreyee Maitra}
\affil{{\small Max Planck Institute For Extraterrestrial Physics, 85748 Garching, Germany}}

\author{Harsha Raichur}
\affil{{\small Nordita, KTH Royal Institute of Technology and Stockholm University, Rosalagstullsbacken, 23, SE-10691 Stockholm}}

\author{Biswajit Paul}
\affil{{\small Raman Research Institute, C V Raman Avenue, Sadashivanagar, Bangalore 560080, India}}
% \email{Email of corresponding author: varun@rri.res.in}
% %\affil{{\small $^{1}$Raman Research Institute, C V Raman Avenue, Sadashivanagar, Bangalore 560080, India}}
% \affil{{\small $^{2}$Department of Astronomy and Astrophysics, Pennsylvania State University, Pennnsylvania, 16802, US}}
% \affil{{\small $^{3}$ St Joseph's College, Singmari, Darjeeling, 734104, India}}
% \affil{{\small $^{4}$Max Planck Institute For Extraterrestrial Physics, 85748 Garching, Germany}}
% \affil{{\small $^{5}$Nordita, KTH Royal Institute of Technology and Stockholm University, Rosalagstullsbacken, 23, SE-10691 Stockholm}}

\begin{abstract}

 We present timing and spectral analysis of data from an observation of the High Mass X-ray Binary pulsar 4U 1907+09 with the \lx\ instrument onboard \as\/. The light curve consisted of a flare at the beginning of the observation, followed by persistent emission. The pulsar continues to spin down, and the pulse profile is found to be double-peaked up to 16 keV with the peaks separated by a phase of $\sim0.45$. Significant energy dependence of the pulse profile is seen with diminishing amplitude of the secondary peak above 16 keV, and increasing amplitude of the main peak upto 40 keV and a sharp decline after that. We confirm earlier detections of the Cyclotron Resonance Scattering  Feature (CRSF) in 4U 1907+09 at $\sim 18.5\pm0.2$ keV in the phase-averaged spectrum with a high detection significance. An intensity resolved spectral analysis of the initial flare in the light curve shows that the CRSF parameters do not change with a change in luminosity by a factor of 2.6. 
 %The \lx\ observation increases the luminosity range over which the CRSF has been detected in 4U 1907+09 without any clear indication of a luminosity dependence. 
 We also performed pulse phase-resolved spectral analysis with ten independent phase bins. The energy and the strength of the cyclotron line show pulse phase dependence that is in agreement with previous measurements. Two features from the current observation: different energy dependence of the two pulse peaks and a strong CRSF only around the secondary peak, both indicate a deviation from a dipole geometry of the magnetic field of the neutron star, or complex beaming pattern from the two poles. 
\end{abstract}

\medskip

\section{Introduction}
Neutron star High-mass X-ray binaries (HMXBs) are binary systems comprising of a neutron star and a massive star orbiting around a common center of mass. \src\\ is such a system discovered in the \emph{Uhuru} survey \citep{giacconi1971}. The companion is O8/O9 Ia supergiant with a mass loss rate of 7$\times10^{-6}$ M$_{\odot}$ yr$^{-1}$ \citep{1980AJ.....85..549S,1989A&A...209..173V,2005A&A...436..661C}, a small fraction of which is accreted by the neutron star. X-ray flux modulation in \src\\ with a stable period of $\sim$8.38 d is attributed to the orbital period of the binary \citep{marshall1980_4u1907}, which is also corroborated from the orbital pulse arrival modulation measurements \citep{intzand1998_4u1907}. The X-ray emission from \src\\ shows clear pulsations with a periodicity of $\sim$437 s \citep{makishima1984}, and it is known to show occasional quasi-periodic oscillations at a frequency of about 65 mHz \citep{intzand1998_4u1907,mukerjee2001}. The pulse profile consists of two peaks with a deep and a shallow minimum in between that are separated in the pulse phase by about 0.5. The energy resolved pulse profiles are almost identical between 2 and 20 keV and above 20 keV the secondary pulse disappears and the shape of the main pulse changes considerably \citep{intzand1998_4u1907}. Though \src\\ was initially found to have a nearly constant spin-down \citep{2001MNRAS.327.1269B}, later observations \citep{2006MNRAS.369.1760B} showed variation in the spin-down rate and multiple torque reversals \citep{2006A&A...458..885F,2009MNRAS.395.1015I}. Period measurements with {\it INTEGRAL} and {\it RXTE} exhibit short term fluctuations in pulse frequency over the long term spin-change rates which are consistent with the random walk model
\citep{2012MNRAS.421.2079S}. A Cyclotron Resonance Scattering Feature (CRSF) at 19 keV was first discovered with \emph{Ginga} \citep{makishima1992,makishima1999} and its first harmonic was detected with BeppoSax \citep{cusumano1998}. The CRSF was further investigated with different instruments like RXTE \citep{2006A&A...458..885F,2009MNRAS.395.1015I}, INTEGRAL \citep{2013ApJ...777...61H} and Suzaku \citep{4u1907_energyrange,maitra2013pulse}.\\

The neutron stars in HMXBs have magnetic fields that are of the order of a few times 10$^{12}$ G. The accreted material is channeled along the magnetic field lines at relatively large distances from the compact object, leading to the formation of extended accretion columns. The exact details of how the magnetic field affects the accretion flow is still a topic of investigation \citep[]{becker_wolf2007,becker2012}.  
It is generally expected that the inflowing material is directed towards the magnetic poles of the neutron star where two hot spots are formed. The gravitational potential energy of the inflowing material is first converted into kinetic energy and then released as X-rays due to shocks and dissipations into the accretion column and on the hot spots \citep{basko1975}. The high magnetic fields of neutron stars can be measured from the cyclotron resonance scattering features. These features are known to be produced as a consequence of cyclotron resonant scattering of X-ray photons in the presence of an intense magnetic field \citep{basko1975}. Formation of the CRSF has been investigated using various techniques \citep{1999ApJ...517..334A,2007A&A...472..353S,2008ApJ...672.1127N,2017A&A...597A...3S}.
The centroid energy of the CRSF is related to the NS magnetic field intensity by the equation: 
\begin{equation}\label{eq:crsfeq}
\crsfeq{}\rm{\,keV} ,
\end{equation}
where $z$ is the gravitational redshift.\\

 The CRSF properties of X-ray pulsars are found to be variable for different reasons \citep{2017JApA...38...50M}. The CRSF line energy evolves with time in sources like Her X-1 \citep{2017A&A...606L..13S} and 4U 1538-522 \citep{2019ApJ...873...62H}. Line energy also shows dependence on X-ray luminosity, for example, in V0332+53 \citep{2017MNRAS.466.2143D}. An unvarying or weakly varying CRSF over a large luminosity range, as seen in 1A 0535+262 \citep{2007A&A...465L..21C,2015ApJ...806..193S} is also of interest. CRSF parameters in most sources show large variation with spin phase, for example, A0535+26, XTE J1946+274, and 4U 1907+09 \citep{maitra2013pulse,4u1907_energyrange}. Accurate measurement of the CRSF also depends on broad band spectral coverage, and it is important to model the underlying continuum well. It is therefore of interest to study the CRSF with different instruments at different epochs, and intensity states of the sources. Pulse phase-averaged and pulse phase-resolved measurements of the CRSF have been carried out previously with AstroSat-LAXPC for 4U 1538--52 \citep{2019MNRAS.484L...1V} and results were found to be consistent with NuStar measurements \citep{2019ApJ...873...62H}.\\

%For most sources, properties of the CRSF are found to be variable, and for different reasons \citep{2017JApA...38...50M}. The cyclotron line energy is known to evolve with time (for example in Her X-1 \citep{2017A&A...606L..13S} ; 4U 1538--52 \citep{2019ApJ...873...62H}), with luminosity (for example in V0332+53 \citep{2017MNRAS.466.2143D}), and with pulse phase \citep{maitra2013pulse,4u1907_energyrange}. An unvarying or weakly varying CRSF over a large luminosity range is also of interest (example 1A 0535+262 \citep{2007A&A...465L..21C,2015ApJ...806..193S}). Accurate measurement of the CRSF also depends on broad band spectral coverage, and it is important to model the underlying continuum well. It is therefore of interest to study the CRSF with different instruments at different times, and intensity states of the sources. Pulse phase averaged and pulse phase resolved measurements of the CRSF has been carried out previously with AstroSat-LAXPC for 4U 1538--52 \citep{2019MNRAS.484L...1V} and was found to be consistent with NuStar measurements \citep{2019ApJ...873...62H}.\\
 
In this paper, we report the results from the timing and spectral analysis of data from \lx\ onboard \as\ for \src\\ with particular emphasis on the CRSF in the phase-averaged and pulse phase-resolved spectra. With a large photon collection area and a moderate energy resolution, the \lx\ detectors onboard \as\ are well suited for the study of CRSF in accretion powered neutron stars. The paper is organized as follows. We give details of the instruments and observations used in \S\ref{sec:obs}. Timing and spectral data reduction analysis and results are given in \S\ref{sec:anres}. Finally, we present a discussion of our findings in \S\ref{sec:dis}. \\

\begin{figure}[t]
 \centering
 \includegraphics[width=0.45\textwidth]{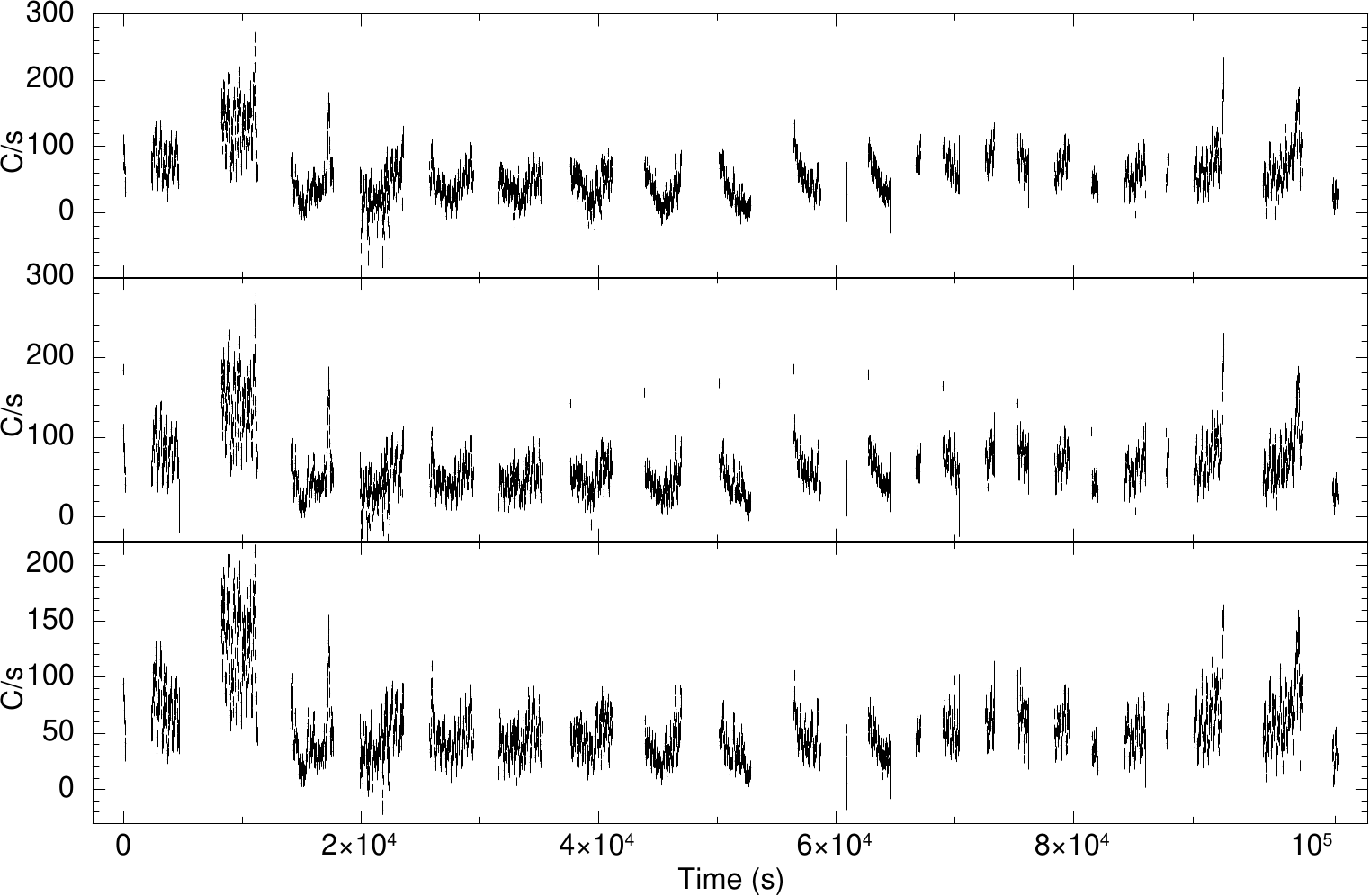}
 \caption{Light curves of \src\\ from the \as\ observation with the \lx. Top, middle and bottom panels show data from the 3 detectors LXP10, LXP20 and LXP30 respectively with a bin size of 10 s. Flaring is seen at the beginning of the observation. }
 \label{lc123}
\end{figure}

\section{Observations}
\label{sec:obs}
 
\as\ was launched on 28th September 2015. The Large Area X-ray Proportional Counter (\lx) is one of the major instruments onboard \as\ which consists of 3 independent proportional counter detectors each with a geometric area of $\sim$ 3600 $\rm{cm}^{2}$ \citep{2017ApJS..231...10A} with slightly different effective areas for each detector and time resolution of 10 $\rm{\mu} s$. Details of the instrument and calibration can be found in \citet{2017ApJS..231...10A}. \src\\ was observed on the 4th and 5th  of June 2017 (Observation ID No 9000001268) for a total of 18 \as\ orbits using the \lx\ instrument in the third cycle of guest observations. This observation was carried out at 0.5 orbital phase away from the peak of the orbital intensity profile. In the Swift-BAT light curve, the source intensity at this orbital phase is $\sim1/3$ of the peak intensity. We have obtained a total of $\sim 47$ ks on-source data from this observation.

\begin{figure*}[t]
\centering
\includegraphics[width=9cm,height=7cm]{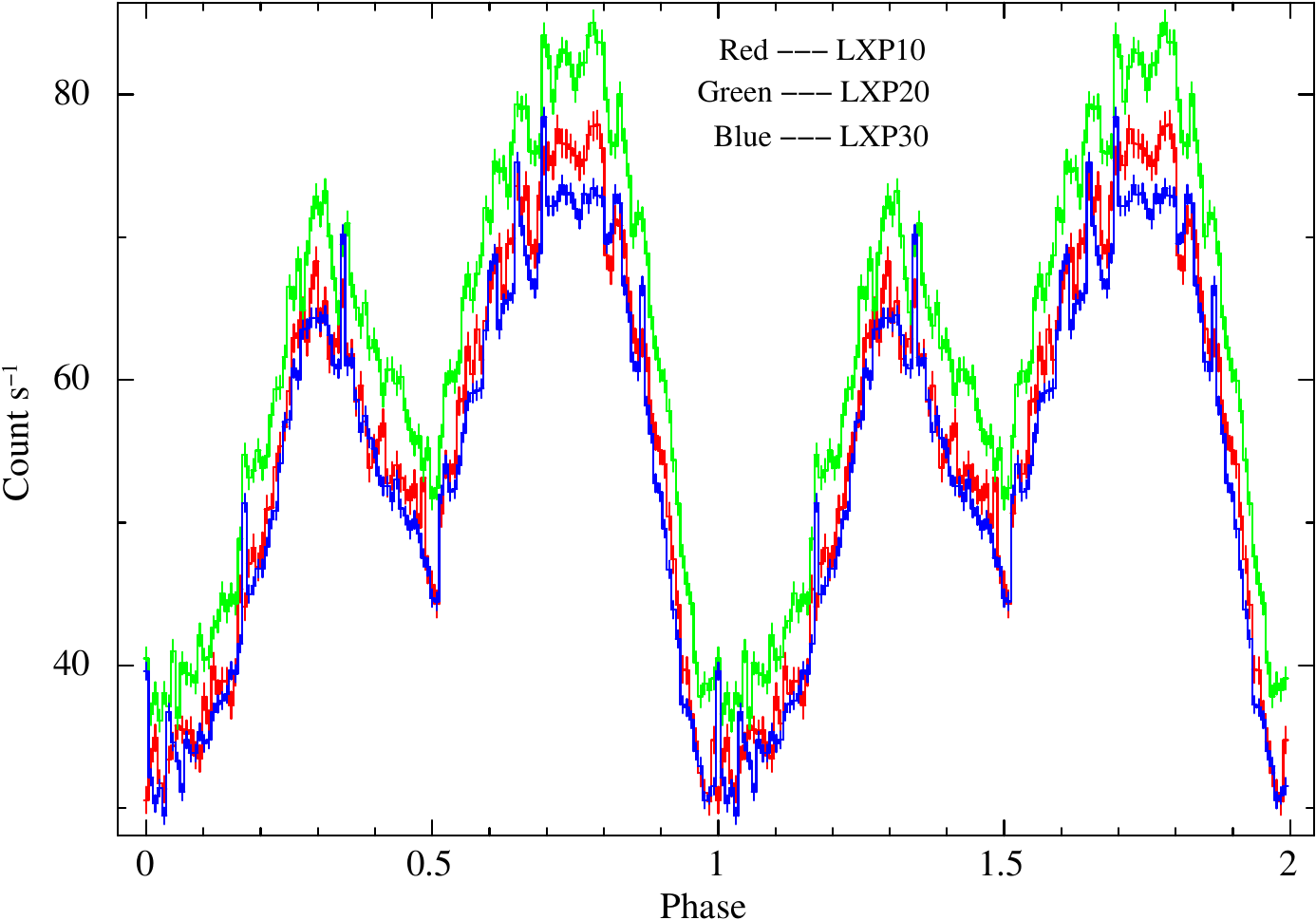}
%\hskip 1cm
\includegraphics[width=7cm,height=7cm]{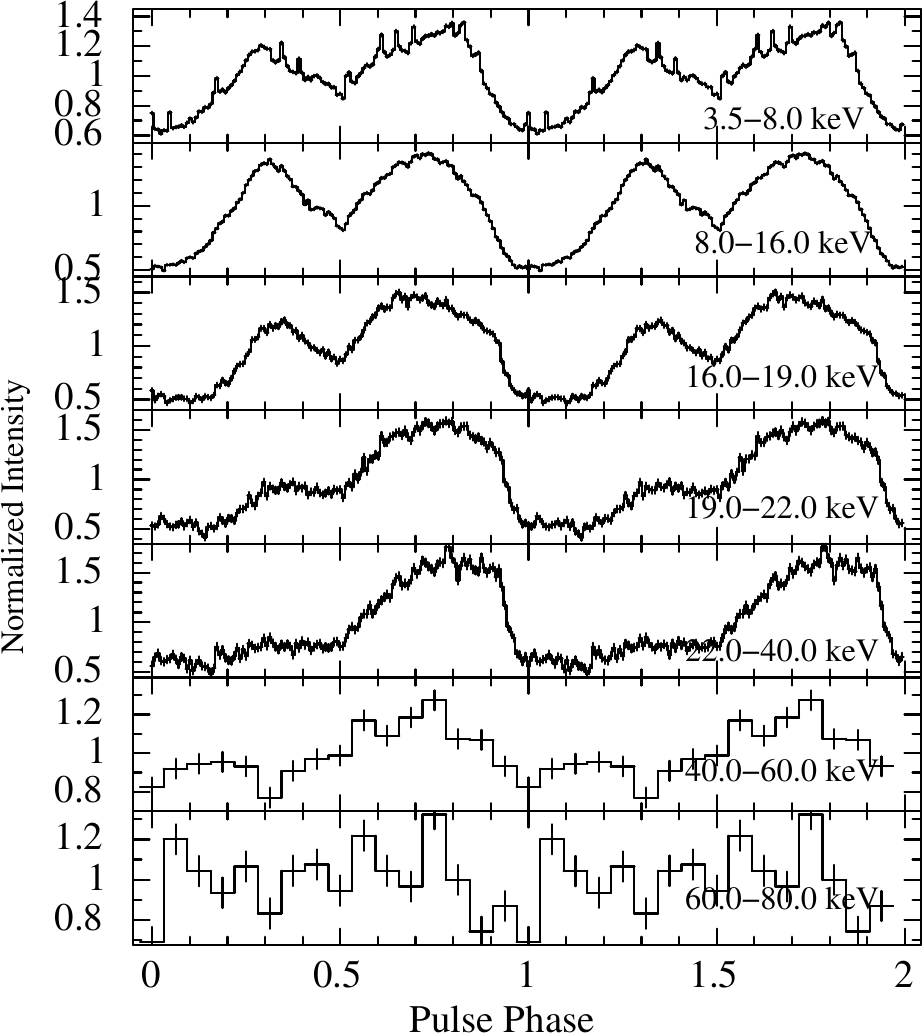}

\caption{Left panel shows pulse profiles from 3 \lx\ detectors in 3.0-80.0 keV energy band plotted with 128 phase bins. Right panel shows energy resolved pulse profiles created from combined data of 3 detectors. 3rd and 4th panels on the right hand side are from 2 bands adjacent to the CRSF Energy.}
\label{ep380}
\end{figure*}

\section{Analysis and Results}
\label{sec:anres}
 We have analyzed data from the Event Analysis (EA) mode of \lx\ from this observation of \src\\. The raw level1 event data files were reduced using \lx\ data analysis pipeline\footnote{\texttt{http://www.rri.res.in/$\sim$rripoc/POC.html}} which produces light curves and spectrum files. Data segments corresponding to the time intervals during earth occultation of source and satellite passage through the South Atlantic Anomaly (SAA) region were removed before the creation of light curves and spectra.

 %For the creation of light curves and spectra, we have removed time intervals corresponding to earth occultation of {\bf the %source} and passage of the satellite through the South Atlantic Anomaly (SAA) region. 

\subsection{Timing Analysis}

A single merged light curve was created with a bin time of 10 ms after eliminating all overlaps between adjacent data downloaded in different orbits. A background light curve was created using time windows during which the source was occulted by the earth. The average background rate was subtracted from the source light curves. The light curves from all three detectors in the 3.0-80.0 keV energy band are shown in Figure \ref{lc123} with a bin size of 10 sec after subtracting an average background count rate for each detector. LXP10, LXP20, and LXP30 have an average source count rate of 54, 61 and 54 per sec respectively. At the beginning of the observation, the source showed a higher count rate for $\sim1/10$ of the observation followed by a lower count rate for the rest of the observation. We carried out a periodicity search on the summed and barycentered light curve with the FTOOL \texttt{efsearch} and obtained a period of 442.33 $\pm$ 0.07 sec. The pulse profiles created using this spin period %and epoch = 57908.00027 (MJD)
from the 3 \lx\ detectors in the energy band 3.0-80.0 keV are shown in the left panel of Figure \ref{ep380}. Pulse profiles from all detectors show an identical double-peaked shape. The first peak is smaller than the second peak. We did not perform orbital correction on the photon arrival time as there are large error bars on most of the orbital parameters of this source \citep{intzand1998_4u1907}. However, we checked whether a period derivative term should be included for the timing analysis. The period search was carried out for a wide range of period derivatives, and it was seen that the data is consistent with zero period derivative. Hence, all subsequent analysis was performed with the above-mentioned pulse period and a zero period derivative.

The emission pattern of many pulsars in the hard X-rays band is different from their emission pattern in soft X-rays which is also affected by absorption. The large effective area of \lx\ allows us to carry out timing analysis in a broad range of 3.0-80.0 keV. In order to investigate the emission pattern at various energies we created energy resolved pulse profiles in 7 energy bands: 3.5-8.0 keV, 8.0-16.0 keV, 16.0-19.0 keV, 19.0-22.0 keV, 22.0-40.0 keV, 40.0-60.0 keV and 60.0-80 keV. Two energy bands 16.0-19.0 keV and 19.0-22.0 keV are chosen in the CRSF energy range of \src\\. The right panel of Figure \ref{ep380} shows energy resolved pulse profiles created with combined data of 3 detectors. Pulse profiles have two peaks with a small peak preceding a higher and broader primary peak. Pulse profiles up to 40 keV are created with 128 phase bins and with 32 phase bins in last two energy bands. Pulsations are detected up to 60 keV.
% and in 60.0-80.0 keV band we don't detect source and hence pulsations are not detected.
The amplitude and shape of the secondary peak have a strong energy dependence with a marked decrease in pulse fraction above the CRSF energy.
%Amplitude of secondary decreases significantly in energy bands around CRSF.
Pulse fraction of the main peak increases from $38\%$ in 3.5-8.0 keV to $58\%$ 19.0-22.0 keV band. In the 22.0-40.0 keV band, it remains the same at $57\%$, and in 40.0-60.0 keV band, it drops to $25\%$.

\subsection{Spectral Analysis}
Spectra were created from all channels which encompass the 3.0-80.0 keV energy band of the \lx\ detectors. Background spectra were extracted from data acquired in time intervals when the source was occulted by the earth. Most of the X-ray photons of energy greater than 34 keV produce double events in \lx\ as an X-ray fluorescence photon emitted from the Xenon atoms is detected in a different detector element. \lx\ detectors have different response matrices depending on the types of events taken to create the spectrum (single or double or all). The response matrix also depends on the gain in the detector, which is variable. 
%can vary over time due to various reasons.
In order to determine this, we first estimated the relative gain of each \lx\ detector during this observation and selected a response matrix accordingly. This was done by creating a spectrum of K-fluorescence photons from the double events. K-events spectrum can be used as a gain calibrator as these photons are monoenergetic\footnote{\texttt{http://www.rri.res.in/$\sim$rripoc/POC.html}}. Once the channels corresponding to the energy of these photons is determined, we compare it with their nominal values determined at the ground to get relative gain for our observation. Significant gain variations were found for LXP20 during our observation. As a result, not all of the double events were registered in the double event spectral window of LAXPCs, which is preset for each detector irrespective of the gain variability.
%Majority of double events in LXP10 are registered in preset window which allowed us to determine its gain.
LXP30 data has large uncertainties in its quantum efficiency and energy response due to gas leakage and thus was not used for spectral analysis. Therefore, for the spectral analysis, only LXP10 data was used. %Data from only LXP10 was used for spectral analysis.
Following \cite{2019MNRAS.484L...1V}, we added a systematic error of 1\% in the energy band of 4.0-10.0 keV and 0.5\% in the band of 10.0-40.0 keV using the tool GRPPHA.\\

\begin{figure*}
 \centering
 \includegraphics[width=0.48\textwidth]{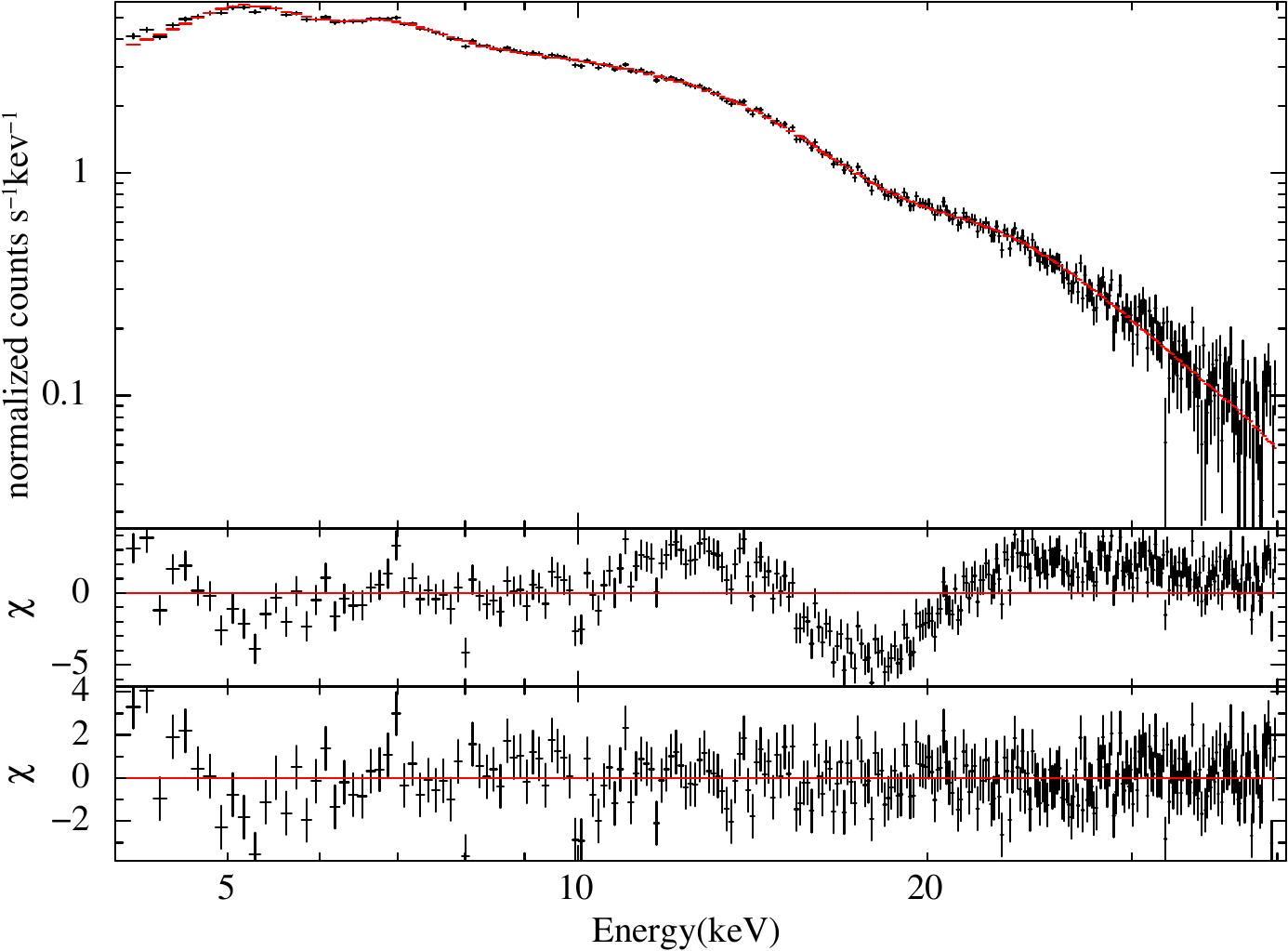}
 \includegraphics[width=0.48\textwidth]{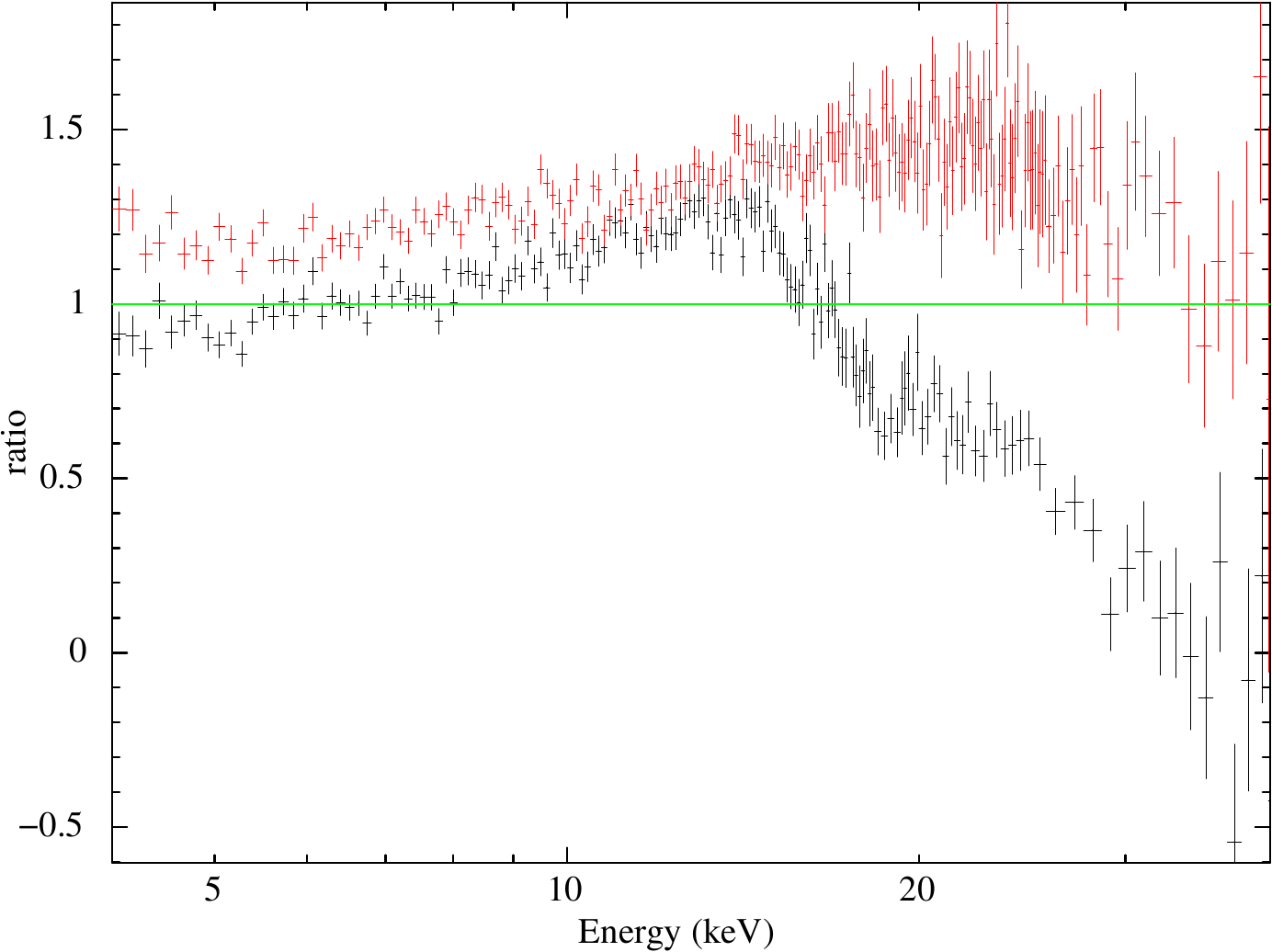}
 \caption{Left panel shows pulse phase-averaged spectrum of \src\\ along with the best fit \neh\ model and residuals to the fit without and with a CRSF component in the model. Right panel shows ratio of two phase resolved spectra with the phase average spectrum. Ratio of spectrum from phase 0.27-0.37 and phase averaged spectra is shown in black, while same is shown for phase 0.59-0.69 in red. These two phases are chosen to demonstrate a high contrast.}
 \label{lx1sp}
\end{figure*}

\subsubsection{Phase-averaged analysis}

In HMXB pulsars, the continuum emission can be interpreted as arising from comptonization  of soft X-rays in the plasma above the neutron star surface. The most commonly used spectral models for HMXB pulsars are high energy cutoff power-law (XSPEC model HIGHECUT), a combination of two negative and positive power laws with exponential cutoff (NPEX), and a thermal Comptonization model (CompTT). Other local models are power-law with Fermi-Dirac cutoff \citep[FDCUT][]{1986bookT} and a smooth high energy cutoff model \citep[\neh][]{2000ApJ...530..429B}. Different continuum models have been used in the past to fit the spectra of \src\\ from different instruments. We fitted the phase-averaged energy spectra with all the five continuum models mentioned above, available as standard or local package in XSPEC \citep{2001ASPC..238..415D}. XSPEC version 12.9.0 was used for spectral fitting.\\

The spectrum was fitted in the energy range of 4.0-40.0 keV. Energy ranges lower and higher than this were neglected due to limited statistics. The background subtracted spectrum was fitted with all the parameters of the continuum model kept free except the absorption column density. %In the \neh\ model, the parameter $\Delta$E was fixed at 5.0 keV \citep{2019MNRAS.484L...1V,2000ApJ...530..429B}.
The column density is not constrained well with \lx\ spectrum above 4.0 keV and was therefore fixed at $1.5\times10^{22}$ atoms $\rm{cm}^{-2}$ \citep{maitra2013pulse}. We have used angr \citep{1989GeCoA..53..197A} table of {\textbf{XSPEC}} for abundances. Iron K$\alpha$ and K$\beta$ lines were detected and modeled as Gaussians with their energy centers fixed at 6.4 keV and 7.1 keV respectively. The line widths were also fixed at 0.1 keV. Separate spectral fits to the continuum with the continuum models mentioned above showed a deep negative feature in the residuals around 19 keV with a width of about 2.5 keV. This is the CRSF feature already known in \src\\ and was modeled with a Gaussian absorption profile (GABS model in XSPEC).
The best fit parameter values and fit statistics for all models fitted to LXP10 spectrum are given in Table \ref{spec_par}. 
%Parameter values and fit statistics {\bf from best-fit spectral model to LXP10 spectrum} are shown in Table \ref{spec_par}.
Among the five continuum models used, the HIGHECUT and the \neh\ model gave comparable fits, and both better compared to the remaining models. To check any correlation between the CRSF and the continuum parameters, we have created contour plots of $\chi^2$ for several combinations of parameters in the HIGHECUT and \neh\, models and no dependencies were found.  
For phase-resolved spectroscopy, we found that the continuum and CRSF parameters in some pulse phases were better constrained with the \neh\ model compared to the HIGHECUT model. Therefore, only \neh\ model is considered for further work in this paper.\\

\neh\ model is a modification of high energy cutoff model, smoothed around the cutoff energy. Analytical form of model \citep{2000ApJ...530..429B} is given by:

\[
  I(E) =
  \begin{cases}
 
                         { \scriptstyle         NE^{-\Gamma}                            }      &        {\scriptstyle    \text{if $E \le E_{cut} - \bigtriangleup E$} }  \\
                         {  \scriptstyle         AE^{3} + BE^{2}+ CE + D                }      &        {\scriptstyle     \text{if $E_{cut} - \bigtriangleup E \le E \le E_{cut} + \bigtriangleup E$} }\\
                          { \scriptstyle         NE^{-\Gamma}exp(\frac{E-E_{cut}}{E_{fold}})  }    &       {\scriptstyle      \text{if $E \ge E_{cut} + \bigtriangleup E$ } }  \\                              
  \end{cases}
\]\\

The constants A, B, C and D are calculated assuming the continuity of the function I(E) and its derivative in the range of $E_{cut} - \bigtriangleup E$ and $E_{cut} + \bigtriangleup E$ \citep{2000ApJ...530..429B}. $E_{cut}$ and $E_{fold}$ are cutoff and folding energies of exponential rollover. The parameter $\bigtriangleup E$ was fixed at 5.0 keV \citep{2019MNRAS.484L...1V}. \\

The left panel of Figure \ref{lx1sp} shows the LXP10 spectrum along with the best fit \neh\ model and the residuals before and after including the CRSF component, thus clearly showing the presence of the absorption feature. Using a CRSF component in the model improves $\chi^{2}$ from 1108 (for 291 degrees of freedom) to 271 for the addition of three parameters.
%We get a probability of chance improvement to be $5\times10^{-39}$ using F-test.
The line is detected with more than 8$\sigma$ significance.
%As can be seen from Table \ref{table}, \neh gives the best fit to continuum of \src\\ with a reduced $\chi^{2}$ of 1.46 for 289 degrees of freedom.
The photon index of $\sim 0.8$, Cutoff energy of $\sim 12$ keV and E-folding energy of about 12 keV are all well within the range for accretion powered high magnetic field pulsars. The CRSF feature is at $\sim18.5$ keV with a width of 2.27 keV and depth of 2.30 keV. The equivalent width of iron K$\alpha$ and K$\beta$ are found to be $\sim220$ eV and $\sim150$ eV respectively.\\

\begin{table*}
\scriptsize
\caption{Best-fit parameters of the pulse phase averaged spectrum of \src\\ from the \as\  observation. The errors correspond to 90\% confidence limits.}
\label{table}
\centering
\begin{tabular}{|p{2.5cm}|p{2cm}|p{2cm}|p{2cm}|p{2cm}|p{2cm}|}
\hline
Parameter 			& \neh\ 		& HIGHECUT 		&  FDCUT  		&NPEX			& CompTT		\\
\hline
$\alpha$ 			& $0.81\pm0.04$ 	& $0.84\pm0.04$		& $0.45\pm0.04$  	&			&			\\
powerlaw$_{norm}^{a}$		& $1.4\pm0.1$		& $1.4\pm0.1$		& $1.06\pm0.09$		&			&			\\
E$_{cut}$ (keV) 		& $12.0\pm0.9$ 		& $11.6\pm0.4$ 		& 12.00	(Frozen)	&			&			\\
E$_{fold}$ (keV) 		& $11.86\pm0.48$ 	& $12.42\pm0.37$ 	& $7.52\pm0.19$ 	&			&			\\
E$_{c}$ 			& $18.5\pm0.2$		& $18.5\pm0.2$		& $19.0\pm0.2$ 		&$19.2\pm0.2$		&$19.0\pm0.2$		\\
$\sigma$ (keV) 			& $2.4\pm0.4$ 		& $2.1\pm0.3$ 		& $2.7\pm0.2$ 		&$3.5\pm0.4$		&$3.0\pm0.2$		\\
Strength 			& $2.48\pm0.56$ 	& $2.09\pm0.30$ 	& $3.71\pm0.34$ 	&$5.7\pm1.1$		&$4.43\pm0.42$		\\ 
NPEX$_{\alpha 1}$		&			&			&			&$0.92\pm0.54$		&			\\
NPEX$_{\alpha_2}$		&			&			&			&-2.0(Frozen)		&			\\
NPEX$E_{cut}$ (keV)		&			&			&			&$4.54\pm0.14$		&			\\
NPEX$_{norm}1^{a}$		& 			&			&			&$3.0\pm2.4$		&			\\
NPEX$_{norm}2^{b}$		&			&			&			&$1.56\pm0.34$		&			\\
CompTT$_{T0}$ (keV)		&			&			&			&			&0.47 (Frozen)		\\
CompTT$_{KT}$ (keV)		&			&			&			&			&$4.89\pm0.06$		\\
CompTT$_{\tau}$(keV)		&			&			&			&			&$18.86\pm0.57$		\\
CompTT$_{norm}^{a}$		&			&			&			&			&$1.44\pm0.03$		\\
$Fe_{K\alpha}$ $Flux^{c}$ 	&$6.6\pm1.5$		&$6.4\pm1.5$		&$6.5\pm1.5$		&$7.4\pm1.7$		&$7.1\pm1.5$		\\
$Fe_{K\alpha}$ eqwidth 	eV	&$216\pm24$		&$209\pm52$		&$208\pm53$		&$254\pm74$		&$231\pm50$		\\
$Fe_{K\beta}$ $Flux^{c}$	&$4.1\pm1.3$		&$4.4\pm1.2$		&$2.5\pm1.2$		&$4.5\pm1.5$		&$3.5\pm1.2$		\\
$Fe_{K\beta}$ eqwidth	eV	&$148\pm45$		&$159\pm45$		&$88\pm42$		&$158\pm50$		&$124\pm42$		\\
Flux$^{d}$ (4-40 keV) 		& $6.21\pm0.06$ 	& $6.21\pm0.04$ 	& $6.19\pm0.05$		&$6.2\pm0.2$ 		& $6.18\pm0.04$		\\
$\chi^{2}_{\nu}$/d.o.f 		& 0.94/288 (3.81/291 	& 0.94/288 (3.58/291  	& 1.14/289 (6.56/292	&1.15/288 (5.19/291	& 1.13/289 (7.08/292	\\ 
				& without \texttt{GABS})& without \texttt{GABS})& without \texttt{GABS})&without \texttt{GABS})	& without \texttt{GABS}	\\
\hline
\end{tabular}
\begin{flushleft}
$^{a}$ x $10^{-2}$ \\
$^{b}$ x $10^{-4}$ \\
$^{c}$ In units of $10^{-4}$ photons cm$^{-2}$ s$^{-1}$ \\
$^{d}$ In units of 10$^{-10}$ erg cm$^{-2}$ s$^{-1}$ \\
\end{flushleft}
\label{spec_par}
\end{table*}

\subsubsection{Pulse phase resolved-spectral analysis}

As mentioned earlier, the pulse profile of \src\\ shows significant energy dependence. Morphology of pulse profile changes from being double-peaked between 3.5-22.0 keV to single peaked beyond the CRSF. Such a sharp change in pulse profile with energy indicates a significant spectral change with pulse phase. To investigate this, we carried out a pulse phase-resolved spectral analysis. This was done by creating spectra in 10 independent phase bins, each of width 0.1. The same background spectrum and response matrix file, as in case of the phase-averaged spectrum were used. The spectra were also fitted in the same energy range as phase-averaged spectrum. The absorption column density and Fe line centers along with their width were frozen as in the case of phase-averaged spectrum. Also, as the width of the CRSF could not be well constrained with the phase-resolved spectra, it was fixed to its phase averaged value. We would like to mention here that since the CRSF energy is later found to be somewhat variable with the pulse phase, line width measured from the phase-averaged spectrum is perhaps broader than the intrinsic line width in phase resolved-spectra.

\begin{figure}[t]
 \centering
 \includegraphics[width=0.52\textwidth]{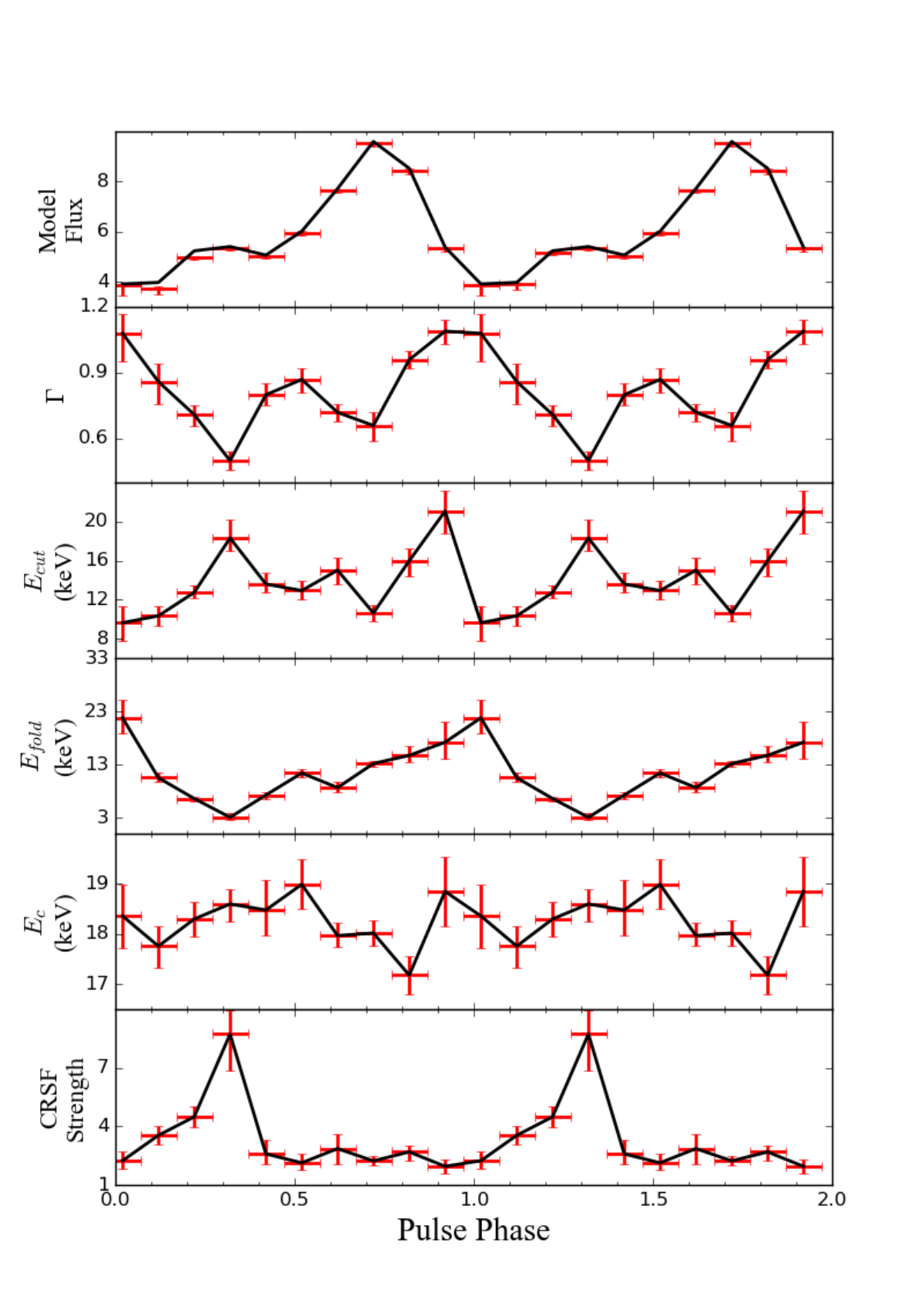}
 \caption{ Variation of spectral parameters with pulse phase is shown here. At top is flux measured in 4.0-40.0 keV band. Next three panels show variations of the continuum components of the \neh\ model and last two panels show the the energy and strength of the CRSF component. In the top panel is flux measured in units of 10$^{-10}$ erg cm$^{-2}$ s$^{-1}$.}
 \label{cop}
\end{figure}

The pulse phase variation of the continuum and the CRSF parameters obtained with the \neh\ model are shown in Figure \ref{cop} with error bars.
The 4.0-40.0 keV flux at different pulse phases is shown in the top panel for comparison with the pulse profile.
The ratio of two phase-resolved spectra with respect to the phase averaged spectrum are shown in the right panel of Figure \ref{lx1sp} highlighting the strong pulse phase dependence of the X-ray spectrum of \src\\. The spectrum at phase 0.59-0.69 is harder compared to the spectrum at phase 0.27-0.37 while the depth of CRSF is smaller in the former. All the Continuum parameters show significant variation with pulse phase. The photon index shows a double-peaked profile that is shifted in phase with respect to the pulse profile. Near the peak of emission profile the photon index decreases indicating a hard spectrum at that phase. Near the minimum of emission profile, the photon index is maximum which means spectrum is comparatively soft in these phases. Hardening of the spectrum at the pulse peak and vice versa is consistent with increasing pulse fraction with energy. $E_{cut}$ and $E_{fold}$ parameters show some variation during most phases, both these parameters peak near the end of the primary peak in the pulse profile. The CRSF parameters also show significant variations with pulse phase. CRSF energy $E_{cyc}$ shows about 12\% variation around the mean value, with it being lowest at the main peak of the pulse. Strength of the CRSF is nearly constant except during the rise of the secondary peak. At phase 0.3, the CRSF strength is larger by a factor of $\sim3$ compared to its value in the phase-averaged spectrum (see right panel of Figure \ref{lx1sp}). 

% Its value being largest at phase 0.3 at which it is larger by a factor of $\sim$3 compared to the same in the phase-averaged spectrum, also evident in the right panel of Figure \ref{lx1sp}.

% We have also investigated whether the pulse phase dependence of the cyclotron line parameters depends on the choice of phase zero. For this, we extracted spectra with overlapping bins of phase width 0.1, each sliding with a phase interval of 0.02 \citep{2019MNRAS.484L...1V}. We point out that the 50 spectra in sliding phase-resolved spectroscopy are not statistically independent. However, the bounds of the spectral parameters obtained from the sliding phase-resolved spectroscopy as shown in \ref{cop} shows that variation of the cyclotron line parameters are insensitive to the choice of phase zero.

\subsubsection{Intensity resolved spectral analysis} 
 We have also performed an intensity dependent spectroscopy from this observation by considering the first two orbit data as high luminosity and rest of observation as low luminosity. We fitted both spectra with the \neh\ model, and the best-fit parameters are given in Table \ref{lxd}. The parameters in high and low luminosity states are in agreement within error bars.

\section{Discussion}
\label{sec:dis}

  In this paper, we have presented the results from timing analysis, pulse phase averaged spectroscopy, intensity resolved spectroscopy, and pulse phase resolved spectroscopy of the HMXB pulsar \src\\ with \as\ \lx. This work is among the early results on the cyclotron lines in accretion-powered pulsars with \as, especially with pulse phase resolved spectroscopy. 4U 1538--522 is the first source for which such studies have been done using \lx\ \citep{2019MNRAS.484L...1V}.

\subsection{Energy dependence of the pulse profile and period evolution}

As is known in \src\\, the \lx\ observations also show strong energy dependence of the pulse profile, the two peaks were varying differently with energy. \emph{Suzaku} observations \citep{4u1907_energyrange} showed a pulse profile with two peaks of comparable amplitude at lower energies and the secondary peak to be much weaker at higher energy. Energy resolved pulse profiles from the \lx\ observation agree with these findings, and we have probed the profile variations further around the cyclotron line energy. We have found that the secondary peak becomes weaker near cyclotron energy (energy bands of 16.0-19.0 keV and 19.0-22.0 keV) and the pulse fraction of primary peak increases from 38 $\%$ in 3.5-8.0 keV band to 58 $\%$ in 22.0-40.0 keV and then it drops to 25 $\%$ in 40.0-60.0 keV range. A change in pulse shape or pulse fraction near the cyclotron line energy has been seen in several pulsars like 1A 1118--61, A 0535+26, XTE J1946+274 \citep{2014EPJWC..6406008M} and V 0332+53 \citep{2006MNRAS.371...19T}. Being at the lower end of luminosity for the persistent HMXB pulsars, a pencil beam pattern is expected in \src\\, with the two pulse peaks being produced from the two magnetic poles. The two pulse peaks are not separated by a phase of 0.5, which indicates either a non-dipole field geometry of the neutron star or a complex beaming profile (a combination of pencil and fan beam geometry) from at least one of the poles. Additionally, the different energy dependence of the two pulse peaks indicates a significant difference in the energy dependence of the beaming pattern in the two magnetic poles in \src\\.

Although we have not carried out photon arrival time correction for the orbital motion, if the nominal values of the orbital parameters determined from the RXTE-PCA observations are used \citep{intzand1998_4u1907}, the current \lx\ observation of \src\\ is done at an orbital phase of 0.18 after T$_{\pi/2}$, corresponding to a line of sight velocity of $-119$ km s$^{-1}$, and a Doppler-shift corrected pulse period of 442.51 seconds. Along with the historical data \citep{2012MNRAS.421.2079S}, it indicates a continued slow down of the pulsar.\\

\begin{table*}[t]
\scriptsize
\caption{ Best-fit parameters obtained from luminosity dependent spectroscopy with the same model as used with phase averaged spectrum. The errors correspond to 90\% confidence limits.}
\label{lxd}
\centering
\begin{tabular}{|p{3cm}|p{3cm}|p{3cm}|}
\hline
Parameter 	 		& High Luminosity 	& Low Luminosity  \\
\hline
$\Gamma$ 	 		& $0.82\pm0.04$ 	& $0.80\pm0.04$  \\
E$_{cut}$ (keV)  		& $11.40\pm0.87$ 	& $13.6\pm1.7$\\
E$_{fold}$ (keV) 		& $13.42\pm0.58$	& $10.2\pm1.2$\\
E$_{cyc}$ 	 		& $18.71\pm0.32$ 	& $18.29\pm0.26$ \\
$\sigma$ (keV) 	 		& $2.71\pm0.54$		& $2.74\pm0.53$\\
Strength 	 		& $2.64\pm0.73$		& $3.3\pm1.3$ \\ 
$Flux^{a}$ (4.0-40.0 keV) 	& $13.4\pm0.1$		& $5.13\pm0.06$ \\

\hline
\end{tabular}
\begin{flushleft}
$^{a}$ In units of 10$^{-10}$ erg cm$^{-2}$ s$^{-1}$ \\
\end{flushleft}

\end{table*}

\subsection{The CRSF and a possible weak luminosity dependence}

 X-ray emission processes in accreting pulsars and the behavior of matter in strong magnetic fields determine the characteristics of their CRSF properties. The CRSF in this source was earlier investigated with several different instruments, most notably with two Suzaku observations which were analyzed differently \citep{4u1907_energyrange, maitra2013pulse, 2014EPJWC..6406008M}. It should be noted that the broad band Suzaku spectrum is obtained with two sets of instruments, XIS (0.7-10 keV) and PIN (15-80 keV), and the relative normalization of the two instruments is often derived from each spectral fit. It is also to be noted that the CRSF feature in \src\\ is close to the energy threshold of PIN detectors. The present measurements are carried out with a single instrument covering a broad band. The low and high energy thresholds of \lx\ are a factor of a few from the CRSF energy, giving a clear detection of the feature with a single instrument. Overall, the CRSF parameters determined from the \lx\ spectrum are in agreement with the previous measurements \citep{cusumano1998, 4u1907_energyrange, maitra2013pulse, 2013ApJ...777...61H}. Different continuum models have been used in the past to fit the spectra of \src\\, and for completeness, we have given results from the spectral fit with all the five models. With three of the spectral models that gave a relatively poorer fit to the \lx\ spectrum, the cyclotron line energy is about 0.5 keV larger than the value obtained with the other two models. We also note that an additional broad 10 keV component that was included in the Suzaku spectrum in \citet{4u1907_energyrange} has been shown to be not necessary in reanalysis of the Suzaku spectrum by \citet{maitra2013pulse}, INTEGRAL spectrum by \citet{2013ApJ...777...61H} and the \lx\ spectrum in the current work.

Some accretion-powered pulsars show a dependence of the cyclotron energy on luminosity. In the super-critical luminosity regime, a negative correlation between CRSF energy and luminosity is expected, and the opposite is expected in the sub-critical regime \citep{becker2012}. Barring singular aspects of some sources, both the expected correlations have been seen in accreting pulsars: a negative correlation in V 0332+53 \citep{2017MNRAS.466.2143D} and a positive correlation in Her X-1 \citep{2007A&A...465L..25S} and GX 304--1 \citep{2012A&A...542L..28K} being prime examples. At much lower luminosity, below L$_{coul}$, the cyclotron line energy is expected to be independent of X-ray luminosity and the same has been observed in A 0535+26 over a luminosity range of two decades and a positive correlation at the highest luminosity range for this source \citep{ 2007A&A...465L..21C, 2015ApJ...806..193S}. Assuming canonical neutron star parameters M = 1.4 $M_{\odot}$ and R = 10 km, \src\\ is well within L$_{coul}$ \citep{2013ApJ...777...61H}. A weak positive correlation between the cyclotron line energy and 5-100 keV X-ray luminosity was reported for \src\\, making it the lowest luminosity cyclotron line source to exhibit this relationship \citep{2013ApJ...777...61H} and indicating that perhaps the sub-critical regime is more relevant for this source. The highest 5-100 keV X-ray luminosity at which CRSF has been reported in \src\\ is about $3\times10^{36}$ erg s$^{-1}$ with the Suzaku observation in 2007. For comparison with the observations mentioned in \citet{2013ApJ...777...61H}, we have extrapolated the flux to the 5-100 keV band and calculated the luminosity for a distance of 5 kpc \citep{2005A&A...436..661C}. The averaged luminosity was $\sim2 \times10^{36}$ erg s$^{-1}$, and the luminosity during the low and high intensity parts as reported in Table 2 was 1.6 $\times10^{36}$ erg s$^{-1}$ and 4.2 $\times10^{36}$ erg s$^{-1}$ respectively, the cyclotron line in all of them being consistent with $18.5\pm 0.2$ keV. Thus the LAXPC observations marginally extend the luminosity range over which the CRSF has been detected in 4U 1907+09 without providing additional support for a luminosity dependence of the line energy. The current observation itself gives a luminosity range of 2.6, over which the CRSF is found to be non-varying beyond 3\%. The CRSF characteristics of \src\\ appear more like A0535+262, with no strong correlation between the CRSF energy and luminosity.

\subsection{Pulse phase dependence of the CRSF}

The \lx\ results are consistent with the earlier reported pulse phase-resolved spectroscopy of \src\\ with Suzaku with six independent phase bins by \citet{4u1907_energyrange} and eight independent phase bins by \citet{maitra2013pulse} respectively. Mainly, the line depth was found to be highest just before the secondary peak and the line energy was found to be highest just after it. \emph{Suzaku} observations were also useful to demonstrate that the pulse phase behavior of CRSF parameters has a similar pattern with a change in luminosity by a factor of 2.5 between 2006 and 2007. The \lx\ pulse phase resolved spectroscopy carried out with 10 independent phase bins shows the line to be strongest around the secondary peak. Given the poor phase resolution in all the phase-resolved spectroscopy with limited statistics, these results are consistent with the Suzaku results. The line being strongest near the secondary peak is somewhat surprising for \src\\, which has a lower luminosity, and  hence is expected to have a pencil beam emission and low depth of CRSF at the pulse peak. This indicates the beaming pattern in the two magnetic poles to be different, a complex beaming pattern, or perhaps a deviation from a dipole geometry of the neutron star magnetic field.  We also note that the variation in cyclotron line energy in \src\\ is about 13\%, which is smaller compared to some other sources like Vela X-1, 1A 1118--61, and XTE J1946+274 \citep{2017JApA...38...50M}.

\section{Conclusions}

In this paper, we have reported results from an observation of the HMXB pulsar \src\\ with the \as\ \lx\ in June 2017. Two peaks of the pulse profile of \src\\ exhibited different energy dependence : the pulsed fraction of the main peak increased till about 40 keV and decreased after that while the secondary peak disappeared at energy above about 20 keV. The pulsar was found to continue on spin-down trend. A CRSF was detected at $18.5\pm0.2$ keV in the X-ray spectrum. A flare was detected at the beginning of the observation with flux 2.6 times compared to the rest of the observation. The cyclotron line energy during the flare was found to be consistent with the rest of the observation. Pulse phase-resolved spectroscopy was carried out with 10 independent phase bins and variations of the CRSF parameters with pulse phase were found to be consistent with previous results from two Suzaku observations of \src\\.

\section{Acknowledgements}

The authors would like to thank the referee for his/her comments and suggestions for improving the manuscript. The research work at Raman Research Institute is funded by the Department of Science and Technology, Government of India. This publication uses data from the \as\ mission of the Indian Space Research Institute (ISRO), archived at the Indian Space Science Data Center (ISSDC). We would like to thank Sreenandini for her help with initial data handling.

\section{Scientific Software Package}

\software{XSPEC, Version 12.9.0 \citep{2001ASPC..238..415D}}

\end{document}